\documentclass[conference]{IEEEtran}
\IEEEoverridecommandlockouts
\usepackage{cite}
\usepackage{amsmath,amssymb,amsfonts}
\usepackage{algorithmic}
\usepackage{graphicx}
\usepackage{textcomp}
\usepackage{xcolor}
\def\BibTeX{{\rm B\kern-.05em{\sc i\kern-.025em b}\kern-.08em
    T\kern-.1667em\lower.7ex\hbox{E}\kern-.125emX}}

 \usepackage{caption}
 \usepackage{lipsum, babel}
  \usepackage{bm} 
 \usepackage{booktabs}
  \usepackage{adjustbox}
  \usepackage{url}
  \usepackage{enumitem} 
  \usepackage{multirow}
\makeatletter
\def\ps@IEEEtitlepagestyle{%
  \def\@oddfoot{\mycopyrightnotice}%
  \def\@oddhead{\hbox{}\@IEEEheaderstyle\leftmark\hfil\thepage}\relax
  \def\@evenhead{\@IEEEheaderstyle\thepage\hfil\leftmark\hbox{}}\relax
  \def\@evenfoot{}%
}
\def\mycopyrightnotice{%
  \begin{minipage}{\textwidth}
  \centering \scriptsize
  Copyright~\copyright~2025 IEEE. Personal use of this material is permitted. Permission from IEEE must be obtained for all other uses. The final version of record is published in the International Conference on Blockchain and Cryptocurrency (ICBC), 2025. The final version of this paper is available on IEEE Xplore. You can access it here: DOI: [10.1109/ICBC64466.2025.11114510]. Available: \url{https://ieeexplore.ieee.org/document/11114510}
  \end{minipage}
}
\makeatother
\begin{document}

\title{Tuning Block Size for Workload Optimization in Consortium Blockchain Networks\\

}

\author{
    \IEEEauthorblockN{Narges Dadkhah, Somayeh Mohammadi, Gerhard Wunder}
    \IEEEauthorblockA{\textit{Department of Mathematics and Computer Science, Freie Universität Berlin, Germany} \\
    \{firstname.lastname\}@fu-berlin.de}
}

\onecolumn 
\maketitle
\begin{abstract}

Determining the optimal block size is crucial for achieving high throughput in blockchain systems. Many studies have focused on tuning various components, such as databases, network bandwidth, and consensus mechanisms. However, the impact of block size on system performance remains a topic of debate, often resulting in divergent views and even leading to new forks in blockchain networks. This research proposes a mathematical model to maximize performance by determining the ideal block size for Hyperledger Fabric, a prominent consortium blockchain. By leveraging machine learning and solving the model with a genetic algorithm, the proposed approach assesses how factors such as block size, transaction size, and network capacity influence the block processing time. The integration of an optimization solver enables precise adjustments to block size configuration before deployment, ensuring improved performance from the outset. This systematic approach aims to balance block processing efficiency, network latency, and system throughput, offering a robust solution to improve blockchain performance across diverse business contexts.
\end{abstract}


\section{Introduction}
Since Nakamoto introduced Bitcoin in 2008 \cite{nakamoto2008bitcoin}, blockchain technology has expanded beyond cryptocurrency into fields such as supply chain, healthcare, and energy. The demand for privacy and diverse business models has increased interest in private blockchains, leading to frameworks such as Hyperledger Fabric (HLF) \cite{androulaki2018hyperledger}, valued for its flexibility and configurable parameters, including block size, which affects performance. Tuning block size is essential for balancing latency and achieving high throughput, measured in transactions per second (TPS) \cite{hyperledger2018}. Smaller blocks enable near real-time transactions but increase processing overhead, while larger blocks reduce overhead but require more block creation time and higher bandwidth, potentially causing bottlenecks. Many organizations rely on default settings due to uncertainty regarding performance impacts. To the best of our knowledge, no mathematical model currently exists for pre-designing network configurations to achieve optimal performance.
To address this challenge, this research introduces a systematic approach to optimize HLF performance by presenting a mathematical model for block size tuning. The model aims to maximize throughput while considering infrastructure constraints, reducing the need for costly hardware upgrades. It is formulated with an objective function, decision variables, and constraints, all represented through mathematical expressions for deeper insight. The model is then solved using a Genetic Algorithm (GA), integrating Machine Learning (ML) predictions for key performance metrics, including block validation and committing times. This approach evaluates the impact of block size, transaction size, and other parameters on throughput, equipping system administrators with a practical tool to optimize HLF configurations.

\section{Transaction Flow in Hyperledger Fabric}\label{TF}
In the first step, clients submit transaction proposals to endorsing nodes. These nodes execute transactions and validate client identities. Endorsing nodes sign the proposals and return them to the client. Once the endorsement policy is met, the client forwards the transaction to the ordering service. Orderers maintain network consensus by collecting transactions, packaging them into blocks, and distributing the blocks to committing nodes. Blocks are released based on configurable conditions, either when a block reaches its maximum size or a specified time interval elapses. Committing nodes validate each transaction within the block and then commit the block. Finally, all nodes update their ledger copies to maintain network consistency.
\section{proposed approach}
\begin{table}[b]
    \caption{Model Parameters/Variables and Their Descriptions}
    \centering
    \begin{adjustbox}{width=\columnwidth, center}
        \begin{tabular}{p{0.28\columnwidth} p{0.65\columnwidth}}
            \hline
            \textbf{Parameter Notation} & \textbf{Description} \\
            \hline
            $TR = \{T_1, T_2, ..., T_n\}$ & Set of transactions \\
            $n$ & Number of transactions \\
            $TR_i$ & $i$-th transaction \\
            $S_i$ & Size of $TR_i$ in bytes \\
            \hline
            $CN = \{node_1, node_2, ..., node_m\}$ & Set of committing nodes \\
            $m$ & Number of committing nodes \\
            $CN_k$ & $k$-th committing node \\
            $BW_k$ & Bandwidth of $CN_k$ \\
            \hline
            $lb$ & Minimum block size limit (bytes) \\
            $ub$ & Maximum number of transactions in a block \\
            $nb = \left\lceil \frac{n}{lb} \right\rceil + 1$ & Number of blocks \\
            $cb$ & Maximum block size limit (bytes) \\
            \hline
                        \hline
            ${VT}_{jk}$ & Validation time of block $j$ by $CN_k$ \\
            ${CT}_{jk}$ & Committing time of block $j$ by $CN_k$ \\
            $y_{ij}$ & 1 if transaction $i$ is in block $j$, otherwise 0 \\
            $t_j$ & Processing time of block $j$ by $CN_k$ \\
            \hline
        \end{tabular}
    \end{adjustbox}
    \label{parameter}
\end{table}
\subsection{The Proposed Mathematical Model}

This research focuses on the process from when the orderer receives transactions until they are added to the ledger. Table~\ref{parameter} outlines the input data and decision variables used in the model.

{\small
\begin{equation}
\min\sum_{j=1}^{nb} t_j \label{obj}\\ 
\end{equation}
}

The objective function \eqref{obj} aims to enhance throughput by minimizing the total processing time for blocks across the network. Here, \( t_j \) represents the total processing time of block \( j \). Minimizing  \( t_j \) increases throughput and helps to determine the optimal block size. The model assumes that the block creation timeout is sufficiently long, making early block cuts due to timeouts negligible.

\begin{equation}
\scalebox{0.75}{$
\begin{aligned}
  &f\left(\sum_{i=1}^{n} y_{ij}, \sum_{i=1}^{n} S_i y_{ij}, BW_k\right) 
  + g\left(\sum_{i=1}^{n} y_{ij}, \sum_{i=1}^{n} S_i y_{ij}, BW_k\right) \leq t_j \\
  &\quad \forall j=1, \ldots, nb, \quad \forall k=1, \ldots, m \label{const1}
\end{aligned}
$}
\end{equation}
{\small
\begin{equation}
  \sum_{i=1}^{n} y_{ij} \leq ub \quad \forall j=1, \ldots, nb  \label{const2}
\end{equation}

\begin{equation}
  \sum_{j=1}^{nb} y_{ij} = 1  \quad \forall i=1, \ldots, n  \label{const3}
\end{equation}

\begin{equation}
 \sum_{i=1}^{n} S_i y_{ij} \leq cb \quad \forall j=1, \ldots, nb \label{const4}
\end{equation}
}
Constraint~\eqref{const1} calculates the block processing time of the $j_{th}$ block. Function \( f \) computes storing time, and function \( g \) represents transaction latency. Storing time consists of Validation Time (\(VT\)), the time required for committing nodes to verify all transactions in the block, and Committing Time (\(CT\)), which is the time needed to add the block to the ledger. The goal is to minimize storing time while keeping latency low, focusing on the impact of block size while assuming other factors remain constant. Both functions depend on three variables: the number of transactions per block \(( \sum_{i=1}^{n} y_{ij} )\), the block size in bytes \(( \sum_{i=1}^{n} S_i y_{ij} )\), and the available bandwidth \(( BW_k )\) for each committing node. ML models predict functions \( f \) and \( g \) using data collected from HLF test networks, where block and transaction sizes were varied under different bandwidth conditions. Prometheus~\cite{prometheus_overview} and Pumba~\cite{pumba2023} were used for monitoring and applying stepped bandwidth limitations. Hyperledger Caliper~\cite{caliper} generated workloads to measure throughput and latency. Several ML models were evaluated using scikit-learn~\cite{pedregosa2011scikit}. Based on the results, XGBoost provided the most accurate validation time predictions, Polynomial Regression was most effective for committing time, favoring lower-degree polynomials for simplicity, and Decision Tree performed best for latency predictions. These models are integrated into the GA, where each solution is represented as a chromosome structured as a matrix, with rows as blocks and columns as transactions. The GA generates an initial diverse population, evaluates fitness using the model’s objective function, and iteratively refines block configurations. Constraint \eqref{const2} limits the number of transactions per block to ensure feasibility. Constraint \eqref{const3} enforces that each transaction is assigned to exactly one block. Constraint \eqref{const4} ensures that the total block size in bytes does not surpass the limit (\( cb \)). \textbf{Remark}: The optimal block size is determined by $\max \left( \sum_{i=1}^{n} y_{ij} \right), \quad \forall j = 1, \ldots, nb$.

\section{Experimental Results and Conclusion}\label{S-result}
\begin{figure}[htbp]
    \centering
    \includegraphics[width=\columnwidth, height= 4.5 cm]{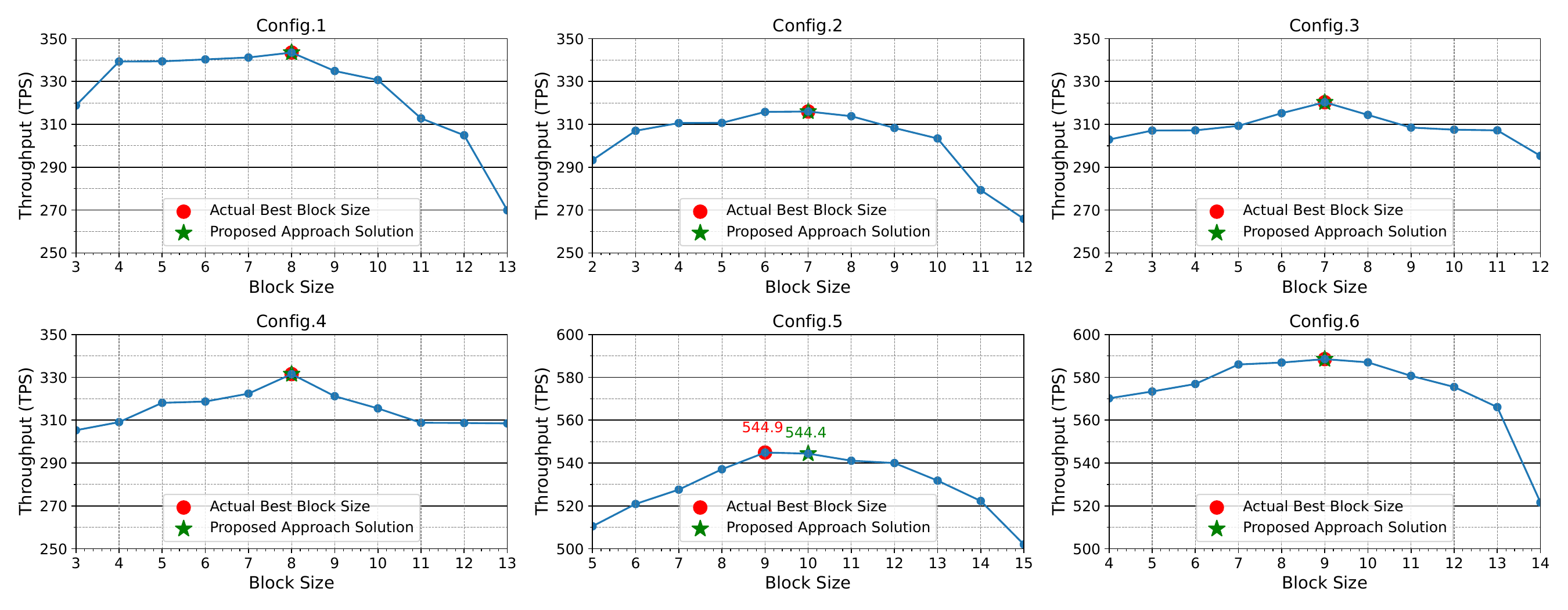}
    \caption{Comparison of the actual best block size and the proposed approach solution}
    \label{Comparison}
\end{figure}

Two experiments were conducted: \textit{Sensitivity Analysis}, to assess sensitivity to key inputs, and \textit{Evaluation of Performance} to evaluate the practical effectiveness of the proposed approach in HLF. The first experiment examined how transaction size, transaction arrival rate, and bandwidth affect block size. In each case, two factors remained fixed while one varied across different ranges, and these data were used in GA. The results showed that as transaction size increases, the optimal block size decreases, adhering to constraint~\eqref{const4}. When the transaction arrival rate increases and reaches a certain value, the block size stabilizes at a specific number of transactions. Notably, the results indicate a slight increase in block size at a specific transaction volume, possibly due to the heuristic nature of the solving strategy, which may select a sub-optimal rather than the global optimal solution. Bandwidth availability significantly influences block size; lower bandwidth requires smaller blocks to avoid congestion, while higher bandwidth supports larger blocks, reducing transmission delays and improving throughput. The second experiment validated whether the suggested block size achieves optimal throughput in HLF. Six configurations were tested by implementing the recommended block sizes and measuring throughput. Fig.~\ref{Comparison} shows six examples of these experiments. The block size was varied around the suggested value to compare performance. Hyperledger Caliper collected throughput data, confirming that the recommended block size consistently achieved the highest throughput. In conclusion, the results demonstrate that, in most cases, the proposed approach can achieve optimal throughput, making it a valuable tool for administrators.

\section*{Acknowledgements}
This work is supported by the Federal Ministry of Education and Research (BMBF) of Germany in the program of “Souverän. Digital. Vernetzt.” through the projects 6G-RIC (16KISK025) and UltraSec (16KIS1682).

\bibliographystyle{ieeetr}
\bibliography{conference_101719}

\begin{thebibliography}{1}

\bibitem{nakamoto2008bitcoin}
S.~Nakamoto, ``{Bitcoin: a peer-to-peer electronic cash system},'' {\em Decentralized business review}, 2008.

\bibitem{androulaki2018hyperledger}
E.~Androulaki, A.~Barger, V.~Bortnikov, C.~Cachin, K.~Christidis, A.~De~Caro, and et~al., ``{Hyperledger Fabric: a distributed operating system for permissioned blockchains},'' in {\em Proceedings of the Thirteenth EuroSys Conference}, EuroSys '18, (New York, USA), Association for Computing Machinery, 2018.

\bibitem{hyperledger2018}
{Hyperledger Performance and Scale Working Group}, ``Hyperledger blockchain performance metrics,'' October 2018.
\newblock V1.01, licensed under Creative Commons Attribution 4.0 International License.

\bibitem{prometheus_overview}
P.~Project, ``Prometheus - monitoring system overview.'' \url{https://prometheus.io/}, 2024.
\newblock [Accessed: 2024-09-20].

\bibitem{pumba2023}
L.~Alexei, ``Pumba: chaos testing tool for docker,'' 2023.
\newblock [Online]. Available: \url{https://github.com/alexei-led/pumba}.

\bibitem{caliper}
H.~Caliper, ``Hyperledger caliper documentation.'' \url{https://hyperledger.github.io/caliper/}, 2023.
\newblock [Accessed: 2024-09-20].

\bibitem{pedregosa2011scikit}
F.~Pedregosa, G.~Varoquaux, A.~Gramfort, V.~Michel, B.~Thirion, O.~Grisel, and et~al., ``Scikit-learn: Machine learning in python,'' {\em Journal of Machine Learning Research}, vol.~12, pp.~2825--2830, 2011.

\end{thebibliography}

\hfill

\end{document}